\documentclass[pre,twocolumn,showpacs,preprintnumbers,amsmath,amssymb,superscriptaddress]{revtex4}

\usepackage{graphicx}

\begin{document}

\title{The shape of jamming arches in two-dimensional deposits of granular materials}

\author{Angel Garcimart\'{\i}n}
\email{angel@fisica.unav.es}
\affiliation{Departamento de F\'{\i}sica, Facultad de Ciencias,
Universidad de Navarra, 31080 Pamplona, Spain.}
\author{Iker Zuriguel}
\affiliation{Departamento de F\'{\i}sica, Facultad de Ciencias,
Universidad de Navarra, 31080 Pamplona, Spain.}
\author{Luis A. Pugnaloni}
\affiliation{Instituto de F\'{\i}sica de L\'{\i}quidos y Sistemas
Biol\'ogicos\\ (CONICET La Plata--UNLP), Casilla de correo 565,
1900 La Plata, Argentina.}
\author{Alvaro Janda}
\affiliation{Departamento de F\'{\i}sica, Facultad de Ciencias,
Universidad de Navarra, 31080 Pamplona, Spain.}

\date{\today}

\begin{abstract}
We present experimental results on the shape of arches that block
the outlet of a two dimensional silo. For a range of outlet sizes,
we measure some properties of the arches such as the number of particles involved, the span, the aspect ratio, and the angles between mutually stabilizing particles. These measurements shed light on the role of frictional tangential forces in arching. In addition, we find that arches tend to adopt an aspect ratio (the quotient between height and half the span) close to one, suggesting an isotropic load. The comparison of the experimental results with data from numerical models of the arches formed in the bulk of a granular column reveals the similarities of both, as well as some limitations in the few existing models.

\end{abstract}


\maketitle

\section{Introduction}
\label{sect:intro}

Granular materials present interesting and unusual physical
properties \cite{duran,jaeger}. One of their most salient features
is the formation of bridges: stable collective structures
comprising several grains that can withstand the weight above them
\cite{mehta1,mehta2,pugnaloni2}. Those bridges, or arches, can cause the jamming of grains in a fixed configuration that is mechanically
stable. It has been argued \cite{cates1} that this is a generic
property of grain assemblies. In granular materials, arches have
been proved to be important for segregation
\cite{duran_segregacion} and non-uniform propagation of forces
\cite{OHern}, as well as directly related to the packing fraction
and the mean coordination number
\cite{pugnaloni1,dorbolo,Nowak}. In particular, it has been
found a clear correlation between the steady state packing
fraction obtained during tapping and the number and size of arches
in the bulk \cite{Pugnaloni3}.

One most common phenomenon that is a consequence of the existence of arches is the arrest
of granular flow at the outlet of a silo. The jamming of a silo has been studied experimentally both in a
two-dimensional contraption \cite{to1,to2,to3,janda1} and in a three-dimensional one \cite{iker1}. In both cases,
attention was paid to the probability of the outpouring being halted by an arch, which depends on the
ratio between the orifice size and the diameter of the grains (if spherical beads are considered). In silo discharge, for intermediate orifice sizes, arches do not arrest the flow; however, they have been shown to be responsible for strong fluctuations of the flow rate \cite{janda2}.

In the last years, numerical simulations have been widely used to approach this problem. In \cite{Pournin} the effect of particle friction and particle size dispersion on arching at the outlet was studied. More recently, a simple probabilistic model has been shown to capture some of the principal aspects of arch formation in hoppers
\cite{Tsukahara}. In other work \cite{Longjas}, a force analysis of the arches formed following simple rules displays good agreement with the jamming probability obtained experimentally. Despite all these advances in the knowledge of silo jamming, the attention payed to the geometry of the arches is scarce. To our knowledge, this issue has only been addressed in the framework of a study based on the Restricted Random Walk Model (RRWM) to explain arch formation in a hopper \cite{to3} where some geometrical features were described, and a comparison with the predictions of the mentioned model was presented. In addition, there is a lack of knowledge about the nature of these blocking arches and their precise relationship with the arches that are developed within a granular column of deposited particles. Intuitively, one could expect that the arches spanning over the orifice are of the same kind as those formed in the bulk, but there is no evidence that this is so.

It is worth mentioning that deciding if a given set of touching grains form an arch is a rather complex problem. In general, the history of the contacts has to be considered \cite{rober}. However, the arch that blocks an orifice is simpler to study since there are no touching grains below the arch. This makes it possible to identify the particles that form the blocking arch from a single snapshot.

In this work, we provide a thorough description of the geometry of
arches that block the outlet of a silo. To this end, we have
implemented a 2D setup which allows for visual inspection of the
arches and a systematic description of their features. This
apparatus is described in the following section. We will also
compare our results with those obtained by numerical simulations
of the arches within the bulk of a granular column \cite{pugnaloni1,rober}.
This comparison can offer important information about the
differences and similarities of arches at the outlet and arches in
the bulk. Some conclusions will be gathered in the last section of
this article.

\section{Experimental setup and methods}
\label{sec:setup}

\begin{figure}
{\includegraphics[width=0.75\columnwidth]{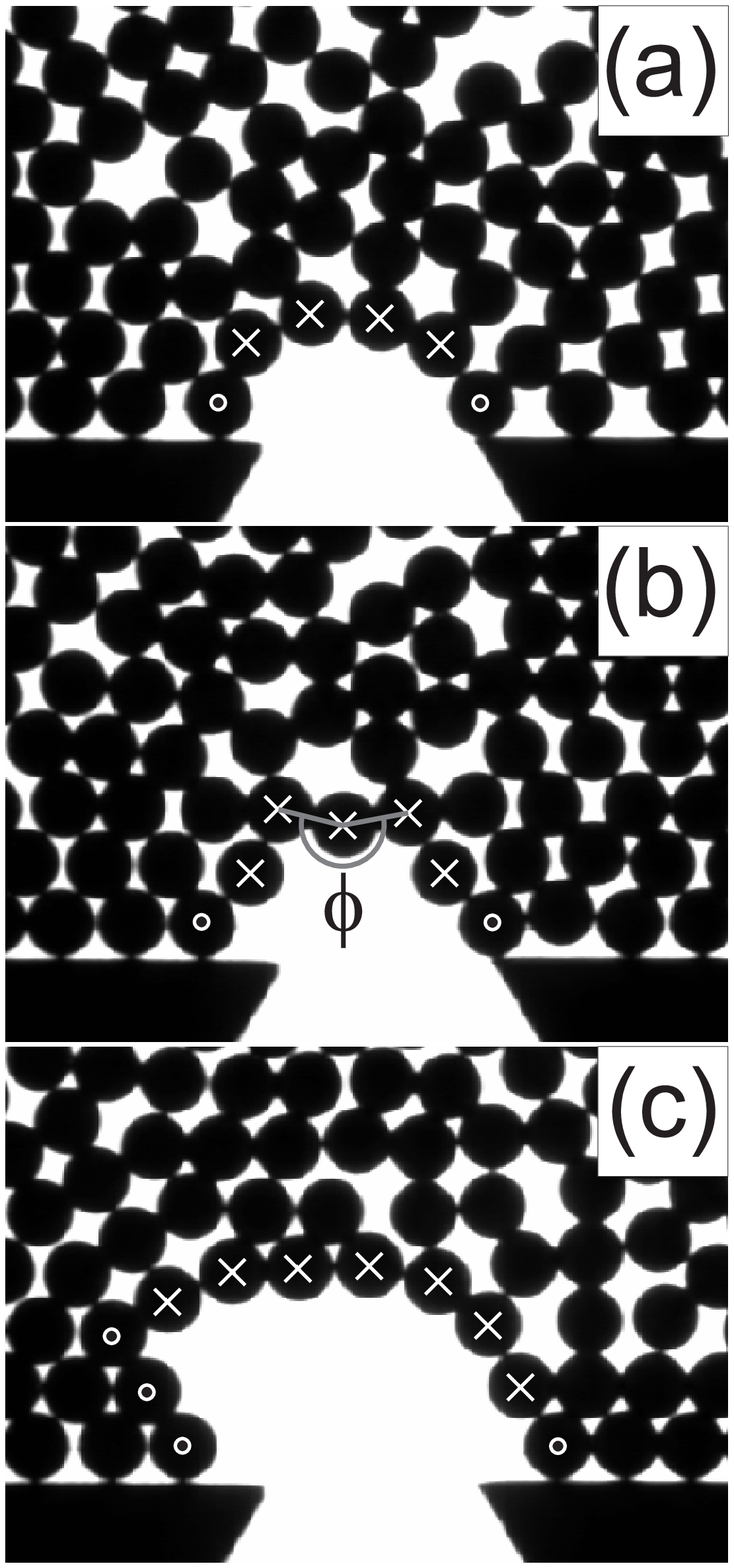}}
{\includegraphics[width=0.9\columnwidth]{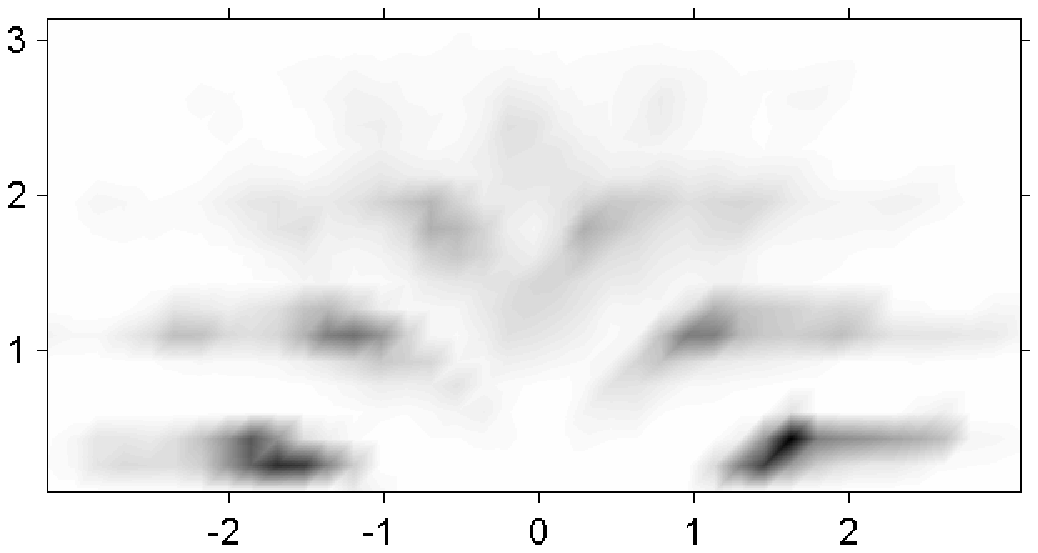}}
\caption{
\label{fig:arco} \textbf{(a)} An arch blocking the outlet of the
silo. White crosses indicate the centers of the particles
that form the arch. The beads forming the base (marked with circles) are excluded from computations, as explained in the text. \textbf{(b)} Example of a bead that hangs from above the equator (the angle $\phi$ is larger than $180^\circ$). \textbf{(c)} In some cases, the arch forms well above the base, with the beads leaning on one or both sides of the banks of a quasi-static funnel made of grains. \textbf{(d)} Frequency map of the position of the beads for about $20000$ arches that blocked the orifice (units are bead diameters; the origin is centered at the exit orifice). Darker areas correspond to places where the probability to find a bead belonging to an arch is higher; beads at the base have been included. All data and photographs in this figure correspond to $R=3.03$.}
\end{figure}

The experimental apparatus consists of a two dimensional rectangular silo (500 mm high and 200 mm wide) made of two glass sheets between which a thin frame of stainless steel is sandwiched. This frame fixes the gap between the two glass sheets. At the bottom, there is an orifice whose size can be changed at will. The granular sample consists of monodisperse stainless steel beads with a diameter of $1.00\pm0.01$ mm. In order to give an idea of the internal friction of the grains, we have measured the avalanche angle, \emph{i. e.} the slope of the plane at which an avalanche develops when a sample is slowly tilted from the horizontal. The value of this angle is $18.5 \pm 0.5 ^\circ$, corresponding to a friction coefficient of about $0.33 \pm 0.02$. The grains are poured along the whole width of the silo from a hopper at the top. The granular deposit within the silo consists of a monolayer of particles because the gap between the glass sheets is slightly larger than the beads. Finite size effects in the
lateral dimensions (corresponding to the width of the recipient) can be neglected as the silo dimensions (width and height) are much bigger than the size of the
beads and the width of the opening at the bottom. A more detailed description of this device can be found in
Ref. \cite{janda1}. The aperture of the exit orifice is given in
units of bead diameters: $R=D/d_0$, where $D$ is the width of the
orifice and $d_0$ is the diameter of the spheres. Since $d_0=1$~mm,
in fact $R$ amounts to the width of the aperture in mm.

Once the silo is filled, grains pour freely from the exit orifice until an arch blocks it. The material flown until a jam occurs is called an avalanche. Under proper lighting
from the back, a still picture (Fig.~\ref{fig:arco}) of the spheres near the orifice is taken (with a standard video
camera) when the flow is stopped and stored for further analysis. Then, the avalanche size is measured ---with a scale at the bottom that collects the grains flown through the orifice--- in
order to validate the measurements with previous results \cite{janda1}. Finally, the flow is restarted by
blowing a jet of compressed air aimed at the orifice. The image and data acquisition system is computerized and
the procedure is automated so that the pictures of many blocking arches can be registered; the limit is imposed by the capacity of the
silo, that has to be refilled whenever the level of grains falls below a fixed threshold of around 300 mm (1.5 times the width of the silo). Seven different outlet sizes were explored, and for each one several
thousand arches were obtained and registered. For some of the
orifices, about twenty thousand arches were obtained in order
to increase the resolution in some calculations.

Even though the particles arrange themselves in a monolayer, a small overlap between the spheres in the images is possible because the gap between the glass panes is necessarily slightly larger than the bead diameter. In our device, however, the overlaps detected are smaller than about $2~\%$ of the particle diameter. We have performed several test runs with a granular sample consisting of round washers, which do not overlap, and we found that the appearance of arches is the same. Spheres were chosen to allow comparison with previous data that can be used as a benchmark \cite{janda1}.

From a digital image such as the one shown in Fig.~\ref{fig:arco}, it is simple to obtain the center of each bead by standard techniques of image processing \cite{russ}. In particular, we have written a code that involves eroding and dilating the image with a disk. Subpixel resolution is attainable, but the main limitation for the accuracy in the measurement of the positions is the overlap of the spheres in the images. We can then easily measure the number of particles $\eta$ in the arch (those marked with an $\times$ in Fig.~\ref{fig:arco}, for instance, $\eta=4$ in Fig.~\ref{fig:arco}(a)), the span (defined as the difference between the horizontal coordinates corresponding to the centers of the leftmost and the rightmost particles marked with white crosses) and the height (the vertical distance between the centers of the highest and lowest particles marked with white crosses).

\section{Arch detection}
The definition of arch as a collection of mutually stabilizing
beads requires the knowledge of which particles sustain each other
in a granular packing. Therefore, the protocols designed to
identify an arch often involve forces between grains or rely on the
history of the grains deposition \cite{rober,Aguirre}, since, in
general, any grain has more contacts than needed to make it
mechanically stable. This information is not easily accessible in
an experiment. This may be one reason for the limited number of experimental studies where the arches are studied \cite{Aguirre,to1,to3}, whereas there is some literature reporting numerical simulations and theoretical studies of arches \cite{pugnaloni2,pugnaloni1,rober,pugnaloni0,Mehta2}. We have
found that much simpler considerations are needed if the focus is shifted to the arches that block the exit orifice, leaving all those
formed in the bulk.

To identify the blocking arch in a given image from our
experiment, we take the first line (bottom up) of touching beads
that span across the orifice (see the beads marked in
Fig.~\ref{fig:arco}(a)). Since our silo has a horizontal bottom,
beads touching the base are stable \textit{per se}; therefore, we
exclude from the arch the two end particles on the base. Note that
if our silo had the shape of a hopper, the particles touching the
bottom should be considered as part of the arch as they would not
be stable \textit{per se}. In many cases, the blocking arch forms
up stream instead of at the outlet. Beads in the silo may develop
a funnel of quasi immobile particles that pile up at both sides of the outlet and arches may form resting
on it [see Fig. \ref{fig:arco}(c)]. In such situations the line of
touching beads detected will bend inwards on one or both sides of
the orifice. If the outermost particles do not touch the base of the silo, we define the blocking arch as the string of particles that span from the leftmost to the rightmost bead in the detected
line of touching beads, excepting those touching the base of the container or forming part of the funnel banks.

\section{Number of particles in the arches}
\label{sec:size}

\begin{figure}
{\includegraphics[width=\columnwidth]{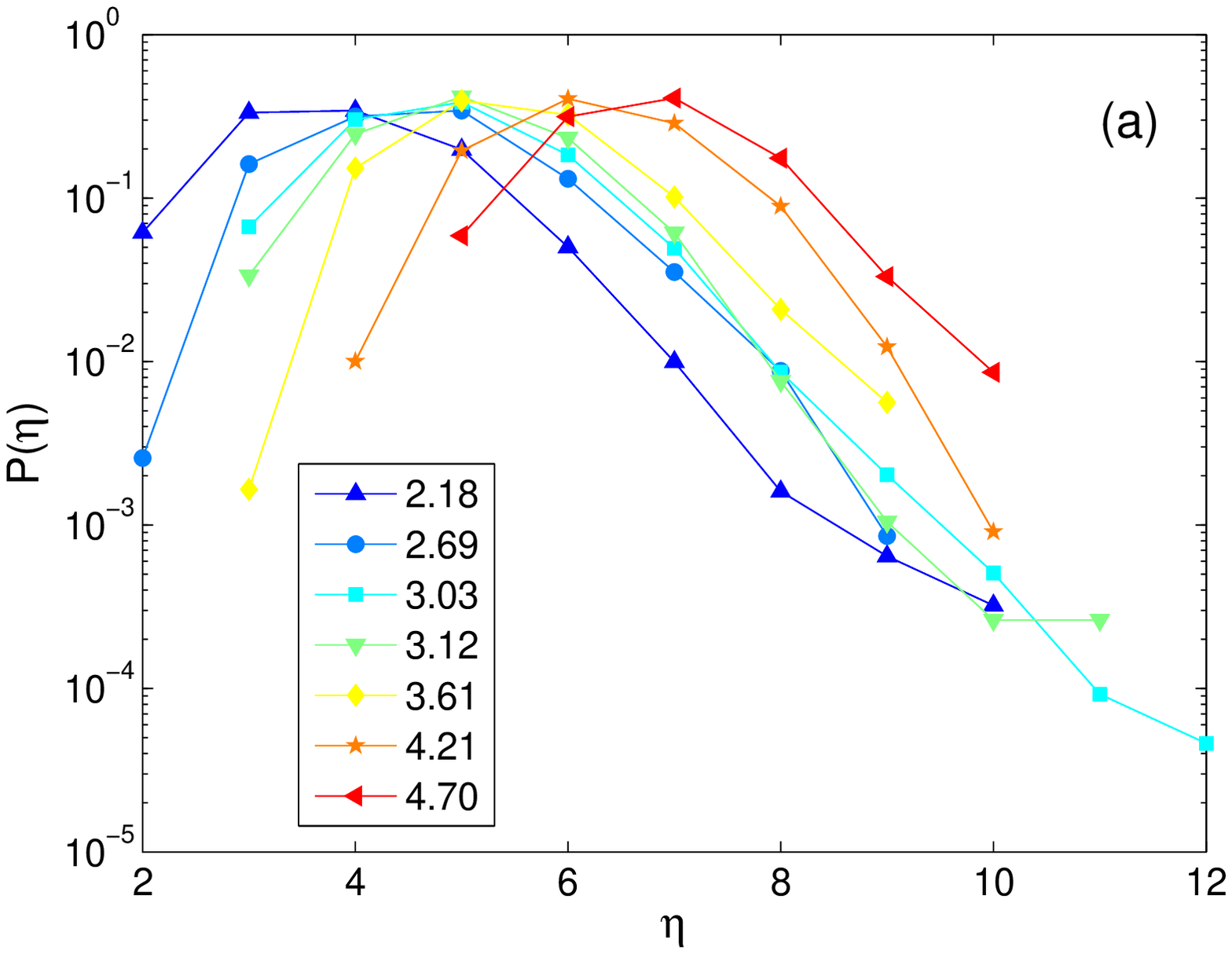}}
{\includegraphics[width=\columnwidth]{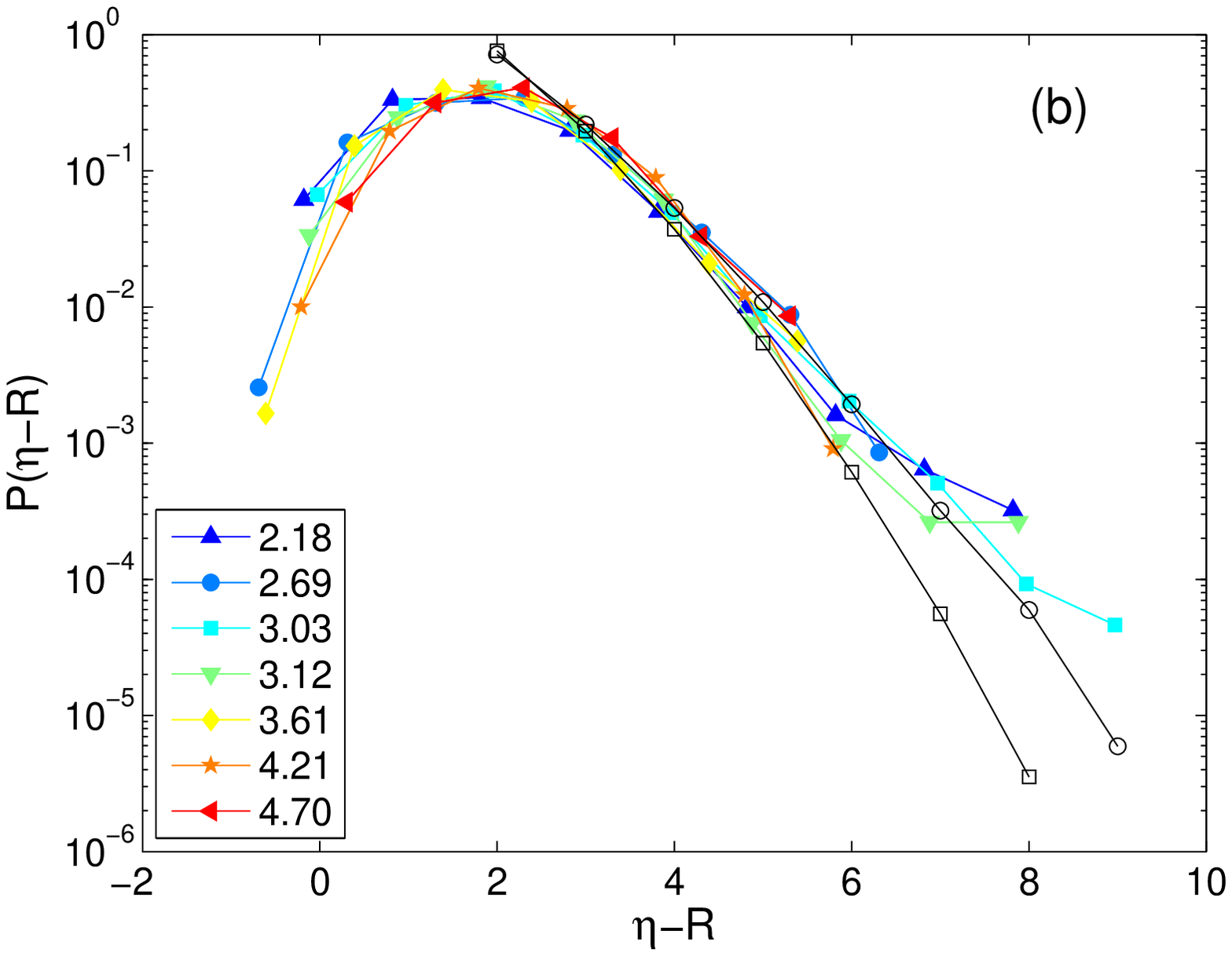}}
\caption{
\label{fig:hnrL} (\textit{Color online})\textbf{(a)} Semilogarithmic plot of the PDF of the number of beads $\eta$ that form the arch for different values of $R$, as indicated in the legend. \textbf{(b)} The same results as in (a), but the variable is now the number of beads minus the
diameter of the orifice. The symbols $\Box$ and $\circ$ display the results obtained respectively from pseudodynamic simulations of hard disks \cite{pugnaloni1} and molecular dynamics simulations of soft disks with friction \cite{rober}.}
\end{figure}

One of the most immediate measurements that can be obtained from the experiment is the number of beads $\eta$ that form each arch. In
Fig.~\ref{fig:hnrL}(a) the probability density function corresponding to different values of $R$ is displayed. (The histograms have been normalized so that the area below the curve is equal to one). Clearly, the number of particles depends strongly on the outlet size, because the bigger the outlet the higher the number of particles that are necessary to form a blocking arch. However, all the distributions display the same trend: an exponential decay for high number of particles and a cutoff for small number of particles imposed by the outlet size. The shape of the number distribution of particles in the arches and the exponential nature of the tails implies that the number of beads per arch has a well defined average for a given orifice $R$.

It is interesting to take a closer look into the tail of the
distribution. Based on a pseudodynamic model, it has been put
forward \cite{pugnaloni1} that the distribution shows a decay
sharper than exponential, such as the series represented with open squares in Fig.~\ref{fig:hnrL}(b). This has been also found in a model of soft disks with static and dynamic friction \cite{rober} (in these simulations, the dynamic and static friction coefficients were set both to 0.5). In probabilistic terms, this would mean that
the price to pay when a new bead is added to an arch would be
higher as the arch grows larger. In order to assess whether this
is the case or not, we have obtained the number of beads for around
$20000$ arches blocking an orifice of size $R=3.03$, which
corresponds to the data set displayed with filled squares in
Fig.~\ref{fig:hnrL}(a). Despite the difficulty of making a
definitive conclusion, the experimental results seem to indicate
that the distribution tail is exponential. It should be noted that
it is impractical to perform more runs because an increase of
about one order of magnitude in the number of arches investigated is needed just to add one data point to the plot. With these data, deviation from a straight line in a logarithmic scale (meaning an exponential decay) is hard to perceive. This would mean that independently of the arch size, the probability of adding one bead to an arch can be calculated multiplying by some fixed factor, as suggested by Mehta et al. \cite{Mehta2}. It should be noted
that in the simulations the distributions depend on the
packing fraction although in all the cases the decay seems sharper
than exponential. The assumption made in some models, such as those cited in \cite{janda1}, that the decay is sharper than exponential, should be reconsidered.

As mentioned above, from the results shown in Fig.
~\ref{fig:hnrL}(a), it is clear that the orifice imposes a cutoff, below which no arch can block the exit. Taking as an
example the results for $R=3.03$, arches with $\eta=4$ seem to be
less abundant than expected due to this fact. This effect is even
more accentuated for $\eta=3$. It may be thought that it is impossible
to find an arch of three particles blocking an orifice of
$R=3.03$. However, as there are usually two beads touching the
silo bottom that are not taken into account when computing $\eta$,
arches with $\eta=3$ are still possible for this value of $R$, although they are scarce.

\begin{figure}
{\includegraphics[width=\columnwidth]{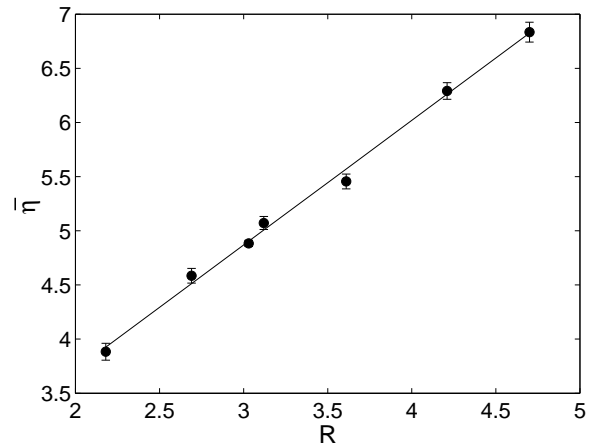}} \caption{ \label{fig:num_R} Average number of beads in the arch that blocks an orifice of size $R$. Error bars correspond to $95~\%$ confidence intervals.}
\end{figure}

We have tried to regroup the PDFs corresponding to the different orifice sizes $R$ by rescaling the variable $\eta$ in various ways. It finally transpires that just by centering all the distribution peaks at the same point regroups all the data sets in a single graph, as displayed on Fig.~\ref{fig:hnrL}(b). This in turn implies that there is a linear relationship between $R$ and the average of the number of beads: $\bar{\eta}=1.41+1.15\;R$ (see Fig.~\ref{fig:num_R}). It should be noted that a linear relationship between the number of beads and $R$ is also obtained if the beads forming the base of the arch (as explained above) are not discarded. In this case, the coefficients of the fit change slightly. We remark here that the range of $R$ explored only covers relatively small orifices; an extrapolation for larger or smaller orifices of this relationship would be unwarranted. Also note that the error bars shown in the plot correspond to confidence intervals; the standard deviation is much larger. This result (\emph{i. e.} a linear relationship between $\eta$ and $R$) is in good agreement with a previous one that had been obtained in a hopper filled with disks for a range of $R$ similar to the one explored here \cite{to3}.

The fact that all the normalized histograms, as presented in Fig.~\ref{fig:hnrL}(b), fall on the same curve, indicates that the nature of the arches that jam the outlet is the same for small and big orifices. In addition, these plots allow a straight comparison of the experimental arches with those obtained within the bulk through numerical simulations. In this latter case $R=0$ has been taken as there is no orifice imposing a cutoff. Despite the difference in the tail explained above, the agreement of numerical results for the PDF with the experimental ones suggests that the arches formed in the orifice (as observed experimentally) are a representative subset of all the arches within the bulk (obtained numerically). Further comments
about this issue are provided in section~\ref{sec:angles}

\section{Geometrical properties of the arches}
\label{sec:span}

\begin{figure}
{\includegraphics[width=\columnwidth]{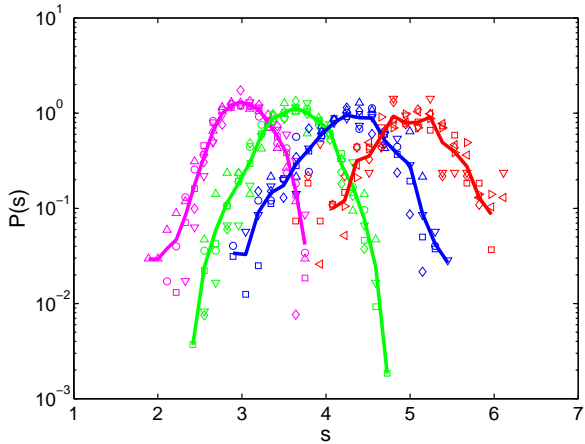}}
\caption{ \label{fig:pdf_env_por_num} (\textit{Color online}) Normalized PDF (in semilogarithmic scale) of the
span ($s$) of the arches formed by $\eta$=5 (\emph{magenta}), 6 (\emph{green}), 7 (\emph{blue}) and 8 (\emph{red}) beads obtained
with different $R$: $\triangle$, 2.18; $\circ$, 2.69; $\Box$, 3.03;  $\triangledown$, 3.12; $\diamond$, 3.61; $\vartriangleright$, 4.21;
$\vartriangleleft$, 4.70. Solid lines are the averages over all $R$ for each $\eta$. }
\end{figure}

\begin{figure*}
{\includegraphics[width=1.9\columnwidth]{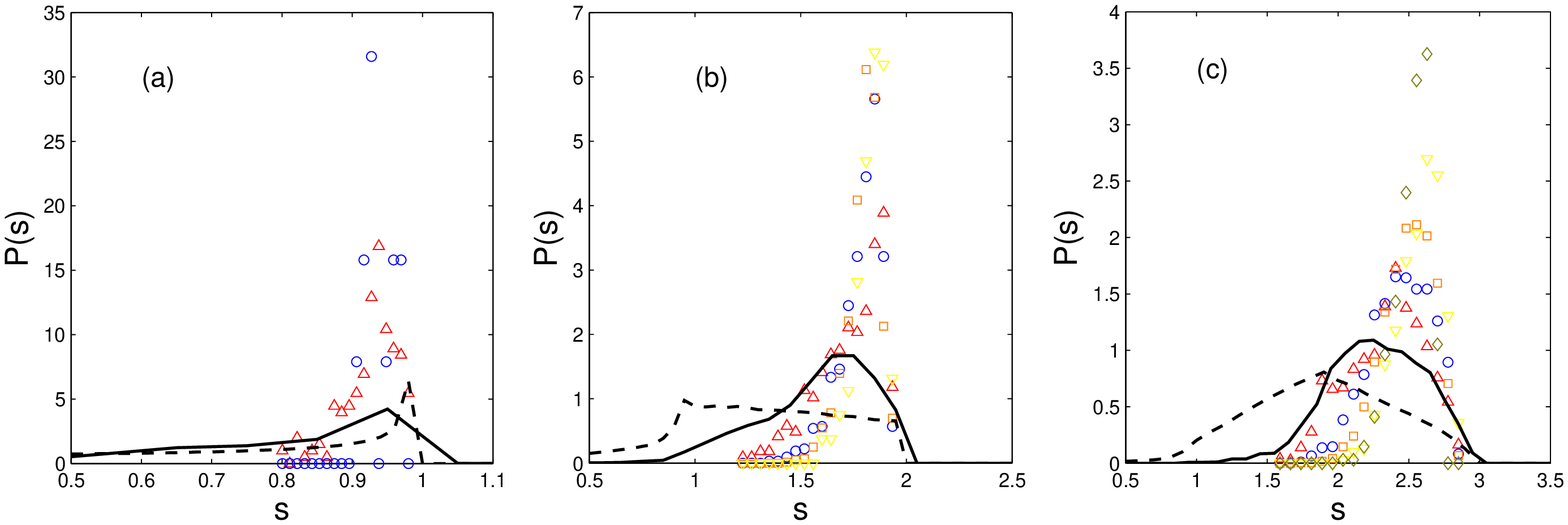}}
\caption{
\label{fig:comp_to} (\textit{Color online}) Normalized PDF of the span, $s$, of the arches formed by \textbf{(a)} 2, \textbf{(b)} 3, and \textbf{(c)} 4 beads obtained with different $R$: $\triangle$, 2.18; $\circ$, 2.69; $\Box$, 3.03;  $\triangledown$, 3.12; $\diamond$, 3.61. Dashed lines correspond to the restricted random walk model proposed in \cite{to1}. Solid lines correspond to arches found in disordered packings of hard disks generated with a pseudodynamic model \cite{pugnaloni1}.}
\end{figure*}

Apart from the number of beads that arches comprise, there are
several features that can be extracted from the experimental observations. One of the most interesting is the horizontal span \cite{to3}. In order to establish a consistent comparison with previous numerical simulations, we have defined the span as the
distance between the abscissae of the centers corresponding to the
two outermost beads of the arch.

One of the first questions that can be asked when looking at the arch span is whether the arches formed with a given number of particles display the same span distribution independently of the size of the outlet that they block. Generally speaking, it is found that this is not the case (Figs.~\ref{fig:pdf_env_por_num} and \ref{fig:comp_to}). But it is true if $\eta > R$ (Fig.~\ref{fig:pdf_env_por_num}), \emph{i.e.} provided that the number of beads is larger than the exit orifice, the arches consisting of a given number of particles display similar features irrespective of the orifice size that they block. In other case, if $\eta \lesssim R$ (Fig.~\ref{fig:comp_to}), differences arise between the span distributions obtained for different $R$, and notably for small spans. This behavior can be understood if one considers that, for a given number of beads, the smaller spans are strongly affected by the cutoff length imposed by the size of the orifice. On the contrary, the arches with larger spans are, in all the cases, larger than the size of the outlet and hence they are not significantly affected by it.

As the arch span is not severely affected by the orifice size (with the condition stated above), it is possible to average the data for arches with the same number of particles, even if they have been obtained with different values of $R$. We present the averages calculated for arches with the same $\eta$ in Fig.~\ref{fig:pdf_env_por_num}. Arches in which $\eta > R$ display distributions that are close to a Gaussian, whereas smaller arches, formed by 2, 3 or 4 particles, display a long tail for small span values (Fig.~\ref{fig:comp_to}). These results reveal clear differences with previous models \cite{to1,pugnaloni1}. Arches of up to 4 particles present larger spans than those predicted by these models. However, the agreement between models and experiment is better for small $R$ (which present broader span distributions) since the influence of the orifice is reduced. Care should be taken when these results are compared with those obtained in a hopper \cite{to1}, where the geometry of the outlet could have a strong influence in the arch properties.

While the span characterizes the size (a direct relationship can be established between $\eta$, the span, and the height of the arches), the shape can be characterized by the aspect ratio: the quotient between half the span and the height of the arch. To be consistent with the definition of span explained above, the height of an arch is defined as the distance between the vertical coordinates of the centers corresponding to the highest and the lowest beads of the arch. If the aspect ratio is equal to one, then half the span is equal to the height. This is obtained, for instance, if the arch is semicircular. The aspect ratio is large for a flat arch, and small for a pointed arch.

\begin{figure}
{\includegraphics[width=\columnwidth]{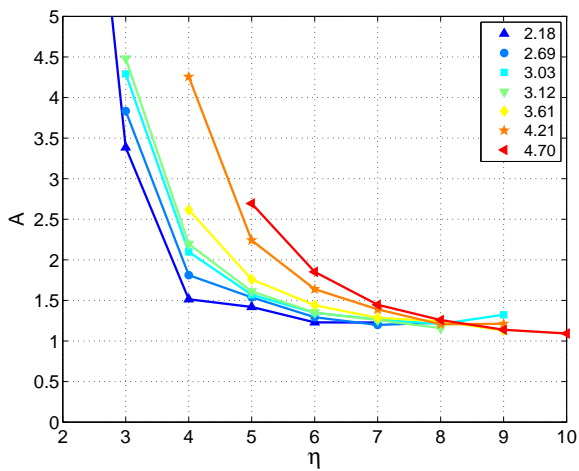}}
\caption{
\label{fig:relaps_num_R} (\textit{Color online})Aspect ratio $A$ versus the number of beads in the arch, obtained for different outlet sizes as indicated in the legend. }
\end{figure}

In Fig.~\ref{fig:relaps_num_R}, the aspect ratio is represented as
a function of the number of beads in the arch for different values
of $R$. The aspect ratio tends to about one when the number of
particles in the arch grows. If the beads at the base of the arch are included, the asymptotic limit is even closer to one. This result supports the hypothesis of semicircular arches, as introduced in \cite{to3}. Although there may be many \textit{defects}, by which we mean beads departing from the angle that would correspond to a perfect semicircle, a circular path can be considered a good starting point, validating the approach used by the cited authors. Additionally, this fact provides a clue about the load sustained by the arches. If the arch is optimized to sustain a vertical load, then it would have an aspect ratio smaller than one. On the contrary, an arch optimized to sustain a horizontal pressure would adopt an aspect ratio bigger than one. A semicircular arch (aspect ratio one) is the preferred shape to optimize an isotropic pressure \cite{ochsendorf,osserman}. Some caution is pertinent here, because a semicircular arch could form for any variety of reasons apart from the optimization of an isotropic uniform load. Besides, even an aspect ratio of one does not guarantee that the arch should be strictly semicircular.

From Fig.~\ref{fig:relaps_num_R}, we can also observe that big arches
have approximately the same aspect ratio irrespective of $R$. This
indicates that the orifice size does not select a particular shape. However, for small arches (small values of $\eta$) the plots for different $R$ display different aspect ratios which seem to diverge when $\eta$ decreases. Indeed, it is clear that the bigger $R$ is, the higher the value of $\eta$ at which the divergence takes place. The reason for this behavior is, again, the cutoff imposed by the orifice. When the number of particles in an arch is similar to the outlet size, the only types of arches that can block the orifice are rather flat and hence the aspect ratio is well above one.

\section{Angles between mutually stabilizing particles}
\label{sec:angles}

\begin{figure}
{\includegraphics[width=\columnwidth]{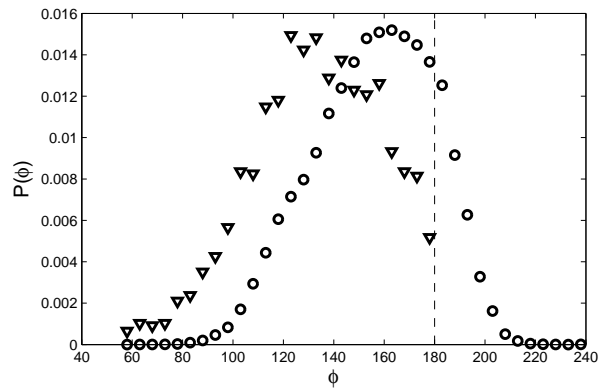}}
\caption{
\label{fig:histang}
PDF for the angles associated to the particles forming an arch,
as shown on Fig.~\ref{fig:arco}(b). Experimental results
(\emph{circles}) and pseudodynamic simulations of hard disks (\emph{triangles}) \cite{pugnaloni1}. The dashed line
indicates $\phi=180^\circ$, the angle at which a particle is
hanging from the equator. Note that in the simulations there are
no beads beyond $\phi=180^\circ$, as they are frictionless.}
\end{figure}

In the previous section, the results of several parameters that
are related with the global shape and size of the arches were presented. In
order to get further insights on the relative position of the
particles within the arches we have measured the angle $\phi$
between the contacts that support each particle in the arch [see
Fig.~\ref{fig:arco}(b)]. Note that every particle in the
arch, except for the end particles, has an associated angle. In
Fig.~\ref{fig:histang} we present the histogram of all the angles
associated to all the beads forming all the measured arches. As we
have not found important differences between the PDFs when taking
into account only large (or small) arches, or when considering
different spans, heights, aspect ratios, or outlet sizes, all the
angles for all the beads in all the cases are displayed combined
in the same plot. The probability distribution of $\phi$ reveals that there is a large number of particles (about $17 \%$ of all registered beads) with an associated angle $\phi$ larger than $180^\circ$. These
cases correspond to beads hanging from above the equator, which
are stable due to static friction since normal contact forces do not contribute to the balance of their weight. From these results, we can conclude that the effect of static friction is quite relevant for arch formation. An example of the important differences that arise when tangential friction is neglected is observed by looking at the distribution of angles between particles obtained from simulations
(Fig.~\ref{fig:histang}). Clearly, the beads do not display angles
larger than $180^\circ$. It must be noted that we have observed the same phenomenon (a considerable number of particles hanging from above
the equator) in test runs with washers. This indicates that
although the gap between the front and rear glass panes is a
little larger than the bead diameters, the small overlapping
that can occur is not the cause for the large number of
particles with $\phi>180^\circ$, because these overlaps are absent if
washers are used.

\begin{figure}
{\includegraphics[width=\columnwidth]{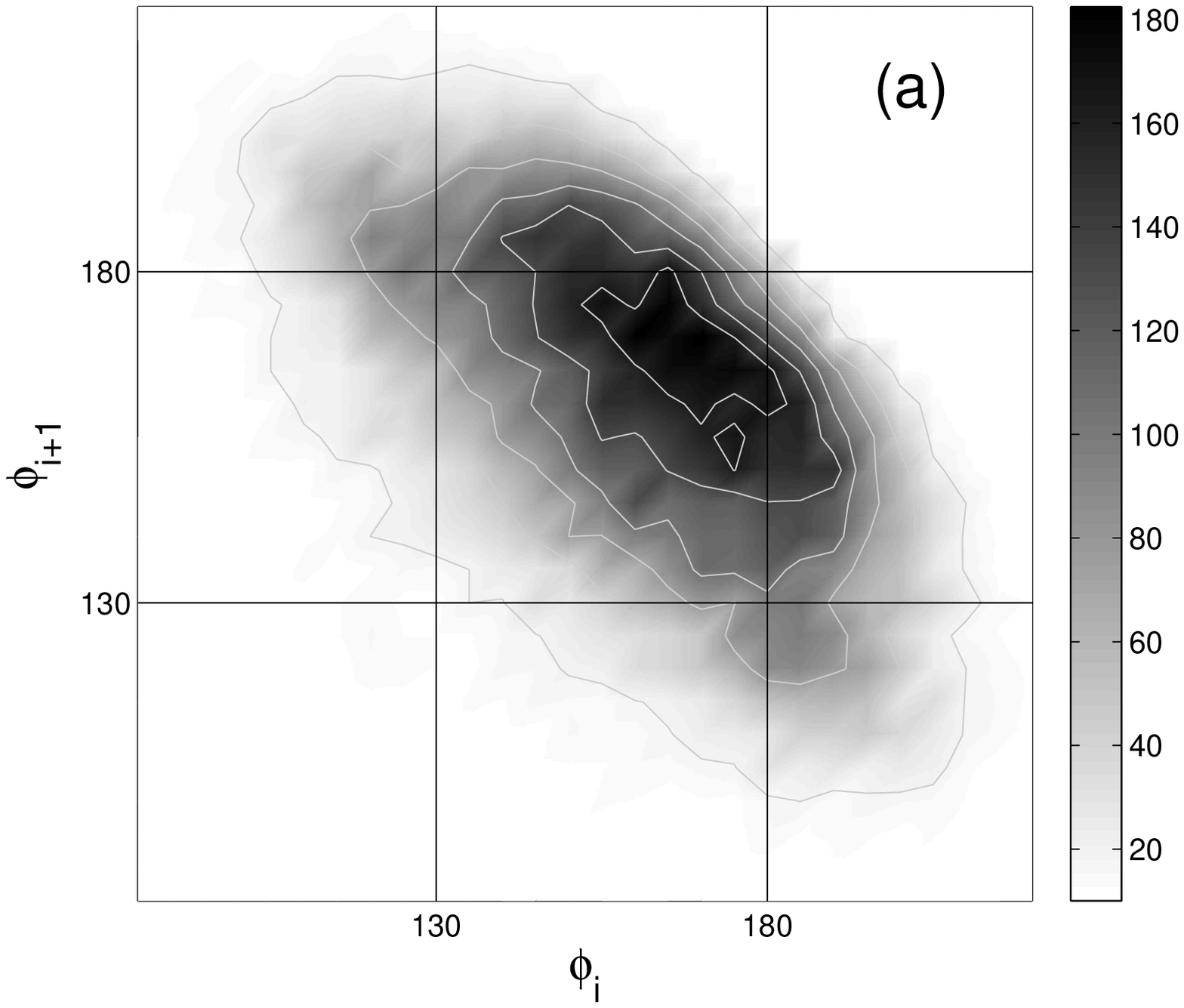}}
{\includegraphics[width=0.9\columnwidth]{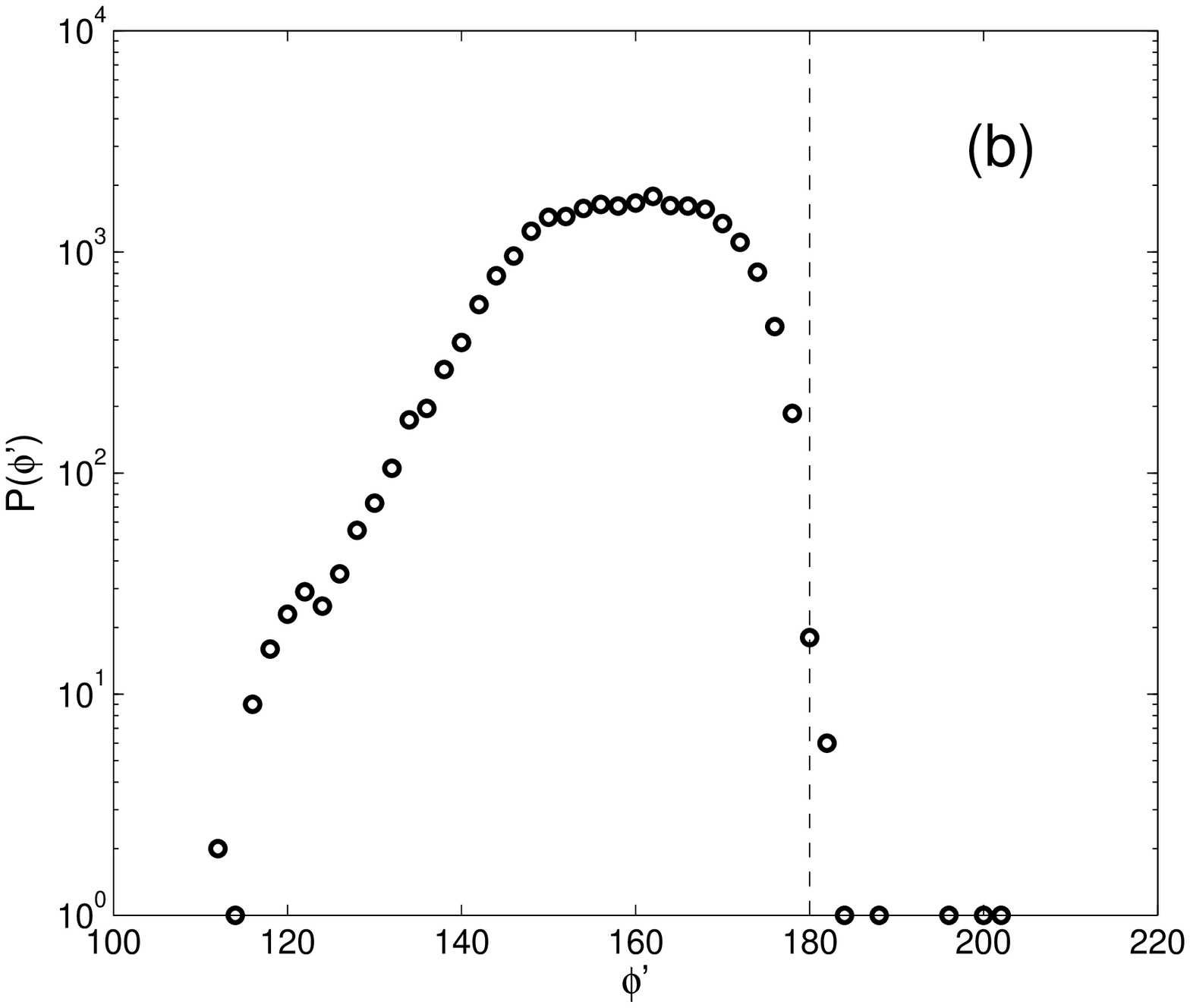}}
\caption{
\label{fig:h1r}
\textbf{(a)}A probability density map showing the angle $\phi$ of the bead $i+1$ versus the angle of the preceding bead $i$ in the arch, where $i$ is the correlative number of the particle in the arch when numbered from left to right. Darker areas indicate higher probability; the probability is not normalized, and the color bar shows the scale.
\textbf{(b)} The PDF for the angle $\phi'=(\phi_i + \phi_{i+1})/2$, which is the average angle between two consecutive links. The dashed line indicates $\phi=180^\circ$. Note the semilogarithmic scale.}
\end{figure}

An interesting question that arises when considering the angles
that form the particles in an arch is whether there is any kind of
correlation among the angles associated to consecutive particles or not. In Fig.~\ref{fig:h1r}(\textbf{a}) we plot the probability density map of the angle of the bead $i+1$ versus the angle of the bead $i$ in the arch. It can be seen that there is a negative correlation between these two quantities. This means that a bead with a large angle will be likely followed by a bead with a small angle, and viceversa. This
result implies that the appearance of two consecutive beads with
large or small associated angles is unlikely. Indeed, it seems
reasonable that the region of $\phi_i>180^\circ$ and
$\phi_{i+1}>180^\circ$ is forbidden as two consecutive beads
hanging from above the equator would conform a very unstable
structure. Additionally, no angles are detected in the region of
$\phi_i<130^\circ$ and $\phi_{i+1}<130^\circ$ as this
configuration would generate an arch with a very small span,
likely unable to block the outlet.

The results displayed in Fig.~\ref{fig:histang} and Fig.~\ref{fig:h1r} (\textbf{a}) seem to be at odds with the constraints of the RRWM. In particular, the existence of angles $\phi>180$ implies that the arches are not necessarily locally convex. Hence, the condition $\theta_i > \theta_{i+1}$ (as defined by authors in \cite{to3}) is not always fulfilled. Moreover, Fig.~\ref{fig:h1r} hints that the angle could be conditioned by the adjacent angles (higher $\phi_i$ means bigger probability of finding a small $\phi_{i+1}$) which is against the hypothesis of random angles proposed in the RRWM. However, by considering these two contradictions together, it results that the RRWM is still valid. The reason for this lies in the fact that a defect is always compensated by the angle associated to its neighbors. Indeed, if we calculate the mean of two consecutive angles $\phi'=(\phi_i + \phi_{i+1})/2$ and we plot its distribution (Fig.~\ref{fig:h1r}(\textbf{b})), we can see that the probability of finding $\phi'>180^\circ$ is so small that it can be deemed statistically insignificant: a clear upper limit in the distribution is displayed at $\phi'=180^\circ$. Then, one could relax the condition stated for the RRWM to $\theta_i > \theta_{i+2}$. Besides, the plateau in the distribution around $\phi'=150^\circ - 170^\circ$ means that there is an approximately constant probability for $\theta$, as required by the model. Then, although the arch may not be convex everywhere due to frictional forces, the RRWM is a good approximation as the local concavity is always compensated by the neighbors.

\section{Conclusions}
\label{sec:conclusions}

In this work we provide a detailed analysis of the geometrical properties of the arches that block the outlet of a 2D silo. The experimental results obtained for different outlet sizes reveal that $R$ does not affect the arch shape provided that the size of the arch is well above the cutoff length imposed by the orifice. We suggest that these arches unaffected by the cutoff imposed by the orifice are representative of the arches formed in the bulk. Besides, we show that, at least for arches with large span when compared to the orifice size, an aspect ratio of one (height equal to half the span) is preferred. This seems to hint that a uniform load (the same from every direction) is at play near the orifice.

The PDF of the number of beads that form an arch displays an exponential tail which implies that the probability of adding a particle to an arch is constant independently of the arch size \cite{Mehta2}. This result contradicts previous simulations where it was suggested a sharper decay than exponential in 2D \cite{pugnaloni1,rober}.

%The other remarkable difference observed between numerical and experimental results is that angles between mutually stabilizing beads in the arch greater than $180^\circ$ are only observed in the experiments. We speculate that those two differences, the distribution of number size and the distribution of angles, are related indeed, and maybe connected to the role of friction. It is plausible that, if angles above $\phi>180^\circ$ are not allowed in simulations, the formation of big arches is prevented as the higher the number of particles in an arch, the higher the probability of finding a particle with an associated angle $\phi>180^\circ$.

Additionally, the measurement of the angles between consecutive
particles in an arch has unveiled the importance of static friction forces between grains in the arch formation process. In particular, we report that in experiments there is a noticeable probability of finding angles greater than $180^\circ$. The existing models of jamming in a hopper assume that the only valid angles $\phi$ between the particles in the arches are smaller than $180^\circ$ \cite{to1}. When friction is taken into account this restriction is fully released \cite{to1}. However, the correlations displayed by the angles of consecutive particles in our experiments may alleviate this limitation of the model. As angles bigger than $180^\circ$ are almost ever compensated by neighbors, the RRWM would likely be valid if \textit{two consecutive angles} are considered.

Note that tangential forces are seldom taken into account in numerical simulations or models of arch formation \cite{Longjas,mehta1,pugnaloni2}. We speculate that the role of static friction must be also of great importance in other granular scenarios. In particular, some models of particle deposition do no consider this effect \cite{pugnaloni2,pugnaloni1,pugnaloni0} and should be revisited.

\begin{acknowledgments}
We thank Diego Maza for his help. We also benefited from discussions with A. Mehta and G. Barker. This work has been financially
supported by Projects FIS2008-06034-C02-01 (Spanish Government),
A/9903/07 (AECI) and PIUNA (Universidad de Navarra). A. J. thanks
Fundaci\'on Ram\'on Areces and Asociaci\'on de Amigos de la Universidad de Navarra for a scholarship. L. A. P.
acknowledges financial support from CONICET (Argentina).
\end{acknowledgments}

\end{document}